\documentclass{ws-ijmpa}
\usepackage{color}
\usepackage[pdftex,plainpages=false,colorlinks=true,citecolor=blue,linkcolor=blue,urlcolor=blue,filecolor=green,bookmarksopen=true]{hyperref}

\usepackage[caption = false]{subfig}
\usepackage{graphicx,epstopdf}
\usepackage{amsmath}
\usepackage{amsthm}
\usepackage{bbm}

\newcommand{\Ocal}{\mathcal{O}}

\newcommand{\1}{\mathbbm{1}}

\newcommand{\Rmath}{\mathbbm{R}}
\newcommand{\Cmath}{\mathbbm{C}}

\newcommand{\Zmath}{\mathbbm{Z}}

\newcommand{\ket}[1]{| #1 \rangle}

\newcommand{\bra}[1]{\langle #1 |}

\begin{document}

\markboth{Alan C. Santos}{Quantum Information Processing via Hamiltonian Inverse Quantum Engineering}

\title{QUANTUM INFORMATION PROCESSING VIA HAMILTONIAN INVERSE QUANTUM ENGINEERING}

\author{ALAN C. SANTOS$^\dagger$\footnotetext{{$^\dagger$}ac\_santos@id.uff.br}}
\address{Instituto de F\'isica, Universidade Federal Fluminense \\ Av. Gal. Milton Tavares de Souza s/n, Gragoat\'a, 24210-346, Niter\'oi, RJ, Brazil}


\maketitle


\begin{abstract}
In this paper we discuss how we can design Hamiltonians to implement quantum algorithms, in particular we focus in Deutsch and Grover algorithms. As main result of this paper, we show how Hamiltonian inverse quantum engineering method allow us to obtain feasible and time-independent Hamiltonians for implementing such algorithms. From our approach for the Deutsch algorithm, different from others techniques, we can provide an alternative approach for implementing such algorithm where no auxiliary qubit and additional resources are required. In addition, by using a single quantum evolution, the Grover algorithm can be achieved with high probability $1-\epsilon^2$, where $\epsilon$ is a very small arbitrary parameter.
\end{abstract}

\keywords{Hamiltonian, quantum algorithms, Deutsch, Grover, Inverse engineering.}


\section{Introduction}

The heart of technologies of the future are based on our ability to control quantum system and designing very small quantum devices. Currently, controlling and protecting quantum systems against decoherence effects is the main challenging task for both theoretical and experimentalists. To protect a quantum system against decohering effects, for example, we can use protocols for speeding up quantum dynamics. In contrast, high speed quantum dynamics requests robust protocols against systematic errors, i.e., uncontrollable deviations in the fields parameters used to drive the system. For this reason, techniques for implementing robust and fast quantum dynamics has woke up interest in recent years. 

For instance, we can consider shortcuts to adiabatic dynamics \cite{Demirplak:03,Demirplak:05,Berry:09} and inverse quantum engineering \cite{Wu:93} as two protocols for speeding up quantum tasks. Hamiltonian inverse quantum engineering (HIQE) is a useful technique to design Hamiltonians able to perform a desired dynamics. In particular, we could highlight the application o HIQE for implementing fast and robust quantum gates necessary for quantum information processing \cite{Santos:18-a}. However, we can find many others interesting applications of both HIQE and shortcuts to adiabaticity techniques, for example in fast transfer/inversion population in nitrogen-vacancy systems \cite{Liang:16}, in Rydberg atoms \cite{Kang:16} and trapped ions \cite{An:16}, as well as applications in two level systems coupled to decohering reservoirs \cite{X-Jing:13,Jing:13,Ruschhaupt:12,Yi:17}, quantum computation \cite{Santos:15,Santos:16-1,Santos:16-2,Santos:18-b}, thermal machines \cite{Funo:17,Adolfo:14,Zheng:16,Abah:17} and others \cite{Torrontegui:13,Huang:17,Chen:16,Chen:18-1,Chen:18-2}.

In this paper we will use HIQE, where no shortcut to adiabaticity is performed, in order to obtain a large class of two-level system Hamiltonians able to drive a quantum system from input state $\ket{\psi_{\mathrm{inp}}}$ to an output one $\ket{\psi_{\mathrm{our}}}$, where $\ket{\psi_{\mathrm{our}}}$ is output of some quantum algorithm (in our case, Deutsch and Grover's algorithm output state). In particular we design Hamiltonians associated to Deutsch and Grover's algorithm. Remarkable we show how HIQE allow us to obtain feasible and time-independent Hamiltonians for implementing such algorithm.

\section{Hamiltonian Inverse Quantum Engineering (HIQE)}

When we start our studies on quantum mechanics, we learn that the dynamics of a quantum system is dictated by Schrödinger equation
\begin{eqnarray}
i\hbar \ket{\dot{\psi}(t)} = H(t)\ket{\psi(t)} \mathrm{ , } \label{SchoEq}
\end{eqnarray}
where $H(t)$ is the Hamiltonian of the system. From this equation, our aim is to solve it in order to find the evolved state $\ket{\psi(t)}$ of the system. Thus, given a Hamiltonian $H(t)$, the problem is to determinate how our system evolves. If we are interested to find a dynamics in particular, obviously we need to solve the above equation for many Hamiltonians until obtaining the desired dynamics. However, sometimes this can be a very hard task, so that we can use HIQE in order to solve this problem.

We can think about HIQE as a method for obtaining Hamiltonians able to drive a quantum system from a input state $\ket{\psi(0)}$ to a target state $\ket{\psi(\tau)}$ through a path $\ket{\psi(t)}$. So, given an evolved state $\ket{\psi(t)}$, we can use HIQE for finding the Hamiltonian $H(t)$ ables to perform this dynamics. In fact, let us write $\ket{\psi(t)} = U(t)\ket{\psi(0)}$, where $U(t)$ is a known unitary quantum operator called \textit{evolution operator}, we can show that the Hamiltonian $H(t)$ associated with $U(t)$ is obtained from equation \cite{Wu:93,Sakurai:Book,Zettili:Book}
\begin{eqnarray}
H(t) = i\hbar \dot{U}(t) U^{\dagger}(t) \mathrm{ . } \label{HamiltGen}
\end{eqnarray}

The operator $U(t)$ has been considered in literature with different proposals. Furthermore, in this paper we are interested in a particular definition of the operator $U(t)$ as discussed in Ref. \cite{Santos:18-a}, where $U(t)$ is written as
\begin{eqnarray}
U(t) = \sum \nolimits _{n} e^{i \varphi_{n}(t)} \ket{\phi_{n}(t)} \bra{\phi_{n}(t)} \mathrm{ , } \label{Uoperator}
\end{eqnarray}
where $\ket{\phi_{n}(t)}$ constitutes an orthonormal bases for the Hilbert space associated with the system and $\varphi_{n}(t)$ are real free parameters. We can see that $U(t)$ satisfies the unitarity condition $U(t) U^{\dagger}(t) = \1$, for any set of parameters $\varphi_{n}(t)$, and it satisfies the initial condition $U(0) = \1$ if we impose initial conditions for the parameters $\varphi_{n}(t)$ given by $\varphi_{n}(0)=2m\pi$ for $m \in \Zmath$. As it was showed in Ref. \cite{Santos:18-a}, from the operator defined in Eq. \eqref{Uoperator} we can find Hamiltonians able to implement quantum gates.

It is important to highlight that we can implement quantum gates from others approaches of HIQE and definitions of the operator $U(t)$. But, as it was discussed in Ref. \cite{Santos:18-a}, these others protocols request physical system with dimension $d\geq4$, two-qubit interaction and auxiliary qubits. For example, a good definition of the operator $U(t)$ has been considered in Ref. \cite{Kang:16}, where additional free parameters can be used for providing experimentally feasible Hamiltonians. However, if we use such method for implement single-quantum gates, for example, we need four-level system. On the other hand, by using the operator in Eq. \eqref{Uoperator}, such gate can be performed in two-level systems. For this reason, we will consider the definition in Eq. \eqref{Uoperator} throughout this paper.

\subsection{Implementing single-qubit quantum gates by HIQE}

Let us consider an arbitrary input state $\ket{\psi(0)} = a\ket{0}+b\ket{1}$, where without less of generality we put $a\in \Rmath$ and $b\in \Cmath$. If we let the system evolves through the operator $U(t)$ from Eq. \eqref{Uoperator}, with $\varphi_{1}(t) = 0$, $\varphi_{2}(t) = \varphi(t)$ and
\begin{subequations} \label{phiState}
	\begin{eqnarray}
\ket{\phi_{1}(t)} &=& \cos [\theta (t)/2] \ket{0} + e^{i\Omega (t)} \sin [\theta (t)/2] \ket{1} \mathrm{ , } \label{phi1} \\
\ket{\phi_{2}(t)} &=& - \sin [\theta (t)/2] \ket{0} + e^{i\Omega (t)} \cos [\theta (t)/2] \ket{1} \mathrm{ , } \label{phi2}
\end{eqnarray} 
\end{subequations} 
with $\theta(t)$ and $\Omega (t)$ being real free parameters, at time $t>0$ the evolved state $\ket{\psi(t)}$ will be given by
\begin{eqnarray}
\ket{\psi (t)} = U_{1}(t) \ket{\psi _{\mathrm{inp}}} = \alpha (t) \ket{0} + \beta (t) \ket{1} \mathrm{ , } \label{psi1Evol}
\end{eqnarray}
where the coefficients $\alpha (t)$ and $\beta (t)$ are given, respectively by
\begin{eqnarray}
\alpha (t) &=& \frac{a \sigma_{+}(t) - \sigma_{-}(t)\tilde{\alpha}(t)}{2} \mathrm{ \ \ \ , \ \ \ } \label{alpha1-beta1}
\beta (t) = \frac{ b \sigma_{+}(t) + \sigma_{-}(t)\tilde{\beta}(t)}{2} \mathrm{ , }
\end{eqnarray}
where we define $\sigma_{\pm}(t)=(e^{i\varphi (t)} \pm 1)$, $\tilde{\alpha}(t) = a\cos \theta(t) + b e^{- i\phi (t)} \sin \theta(t)$ and $\tilde{\beta}(t) =b\cos \theta(t) - a e^{i\phi (t)} \sin \theta(t)$. Thus, we can associate the parameters $\theta(t)$, $\varphi (t)$ and $\Omega (t)$ with an arbitrary rotation of a single-qubit state in Bloch sphere \cite{Santos:18-a}, i.e., an arbitrary quantum gate.

The Hamiltonian that evolves the system as in Eq. \eqref{psi1Evol} is obtained from Eq. \eqref{HamiltGen} and it can be written as
\begin{eqnarray}
H_{1}(t) = \frac{1}{2} \left[ \omega_{x}(t) \sigma_{x} + \omega_{y}(t)\sigma_{y} + \omega_{z}(t)\sigma_{z} \right] \mathrm{ , } \label{GenericH}
\end{eqnarray}
where
\begin{subequations}
\begin{eqnarray}
\omega_{x}(t) &=& ( \cos \varphi -1 ) \dot{ \Omega } \cos \Omega \cos \theta \sin \theta 
+ ( \dot{\theta} \cos \theta \sin \varphi + \dot{\varphi} \sin \theta )\cos \Omega
\nonumber \\
&+& [  \dot{\Omega} \sin \theta \sin \varphi + ( \cos \varphi -1 ) \dot{\theta} ] \sin \Omega \mathrm{ , } \label{omegaX} \\
\omega_{y}(t) &=& ( \cos \varphi -1 ) \dot{ \Omega } \sin \Omega \sin \theta \cos \theta 
+ \sin \Omega ( \dot{\theta} \cos \theta \sin \varphi + \dot{\varphi} \sin \theta ) 
\nonumber \\
&+& [ \dot{\Omega} \sin \theta \sin \varphi - ( \cos \varphi -1 ) \dot{\theta} ] \cos \Omega \mathrm{ , } \label{omegaY} \\
\omega_{z}(t) &=& - \dot{\theta} \sin \theta \sin \varphi  -( \cos \varphi -1 ) \dot{\Omega} \sin ^{2}\theta  + \dot{\varphi} \cos \theta \label{omegaZ} \mathrm{ . }
\end{eqnarray}
\end{subequations}
In general, there are systems in which the $y$-component of the Hamiltonian in Eq. \eqref{GenericH} can be hard to implement. For instance, in systems composed by Bose-Einstein condensates in optical lattices \cite{Bason:12} and some superconducting circuits \cite{Johnson:11,Harris:10,Orlando:99,You:05}. Remarkably, by using our approach we can choose the parameters $\theta(t)$, $\varphi (t)$ and $\Omega (t)$ so that $\omega_{y}(t)=0$. In fact, without loss generality we can put $\Omega (t) = 0$ and $\theta(t) = \theta_{0}=$ cte, so that $\omega_{x}(t) = \sin (\theta_{0}) \dot{\varphi}(t)$, $\omega_{y}(t) = 0$ and $
\omega_{z}(t) = \cos (\theta_{0}) \dot{\varphi}(t) $. In conclusion, an arbitrary single-qubit gate can be implemented without two qubit interaction and no additional resource. By using concrete examples (algorithm), in the next sections we will show how we can adequately choose these parameters.

\begin{figure}
\centering
\subfloat[Deutsch's circuit]{\includegraphics[scale=0.33]{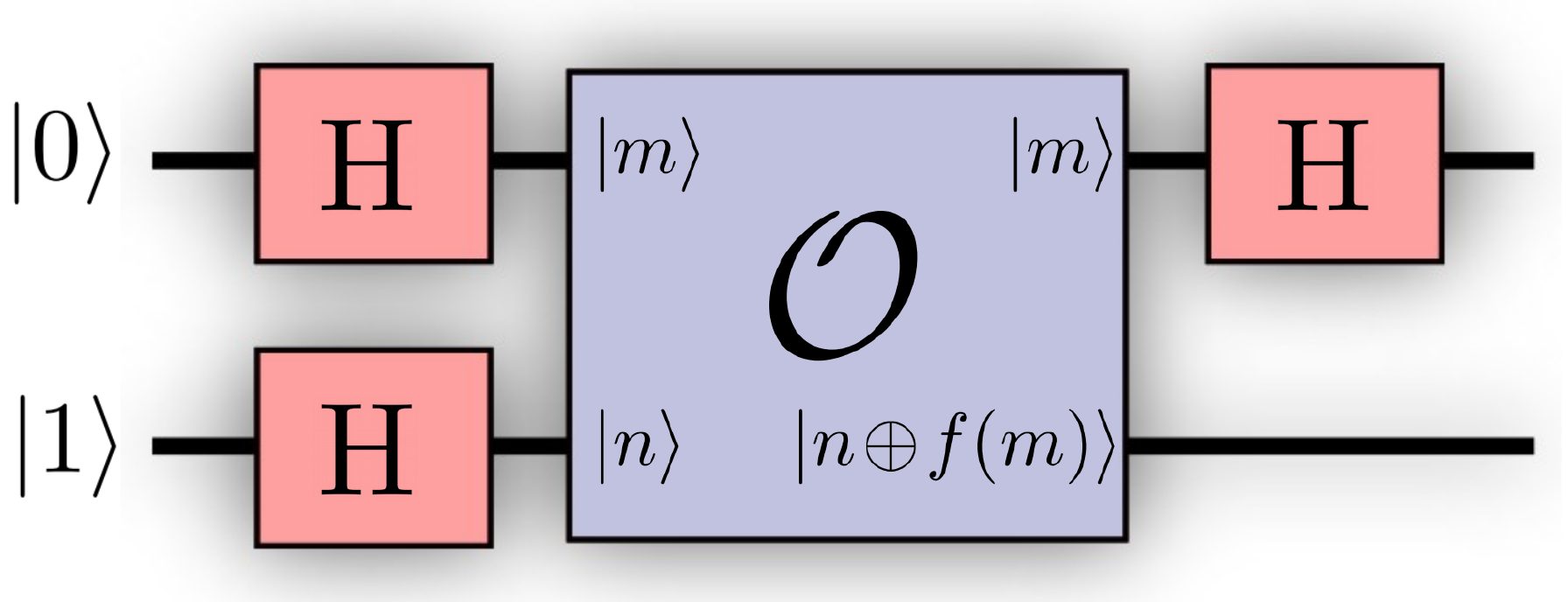} \label{DeutschCirc}}\quad
\subfloat[Alternative to Deutsch's circuit]{\includegraphics[scale=0.32]{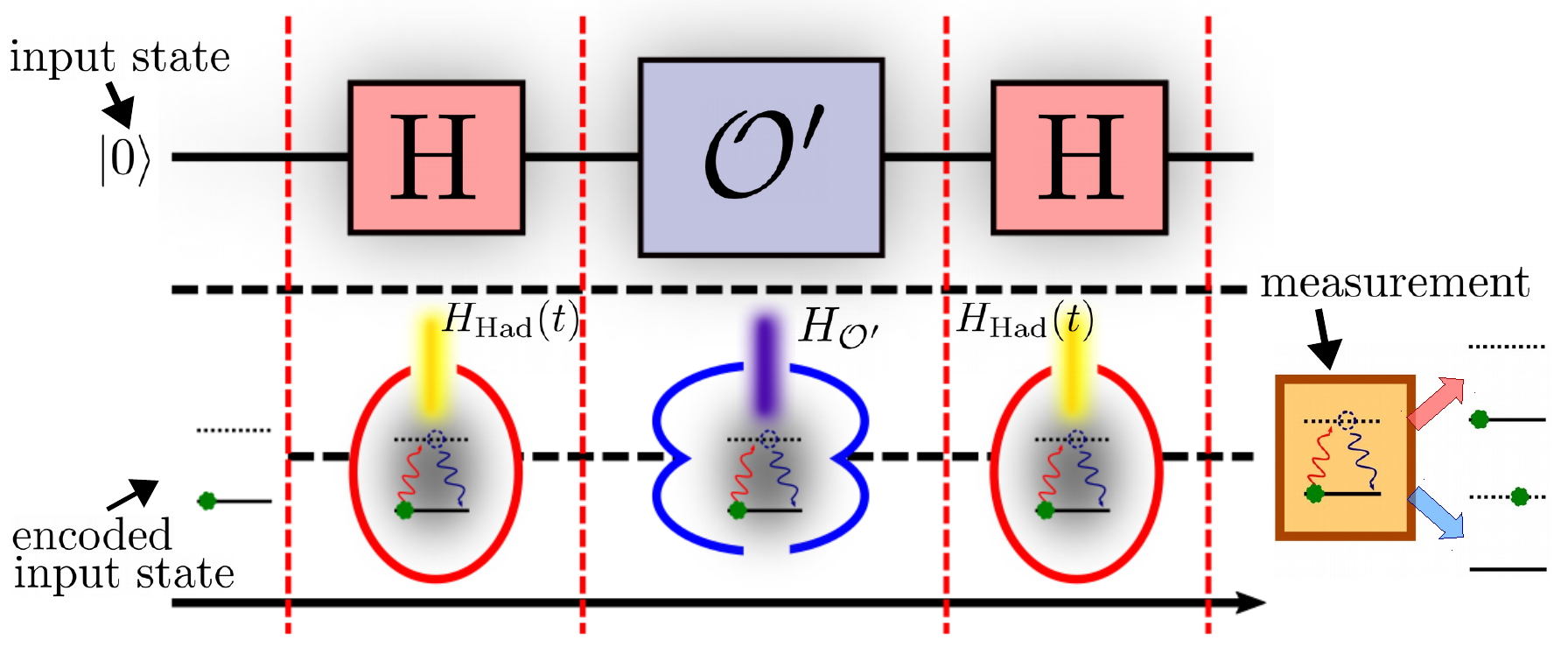}\label{AlternDeutschCirc}}
\caption{({\color{blue}a}) Schematic representation of the Deutsch's circuit. ({\color{blue}b}) Circuit and schematic representation of two-level system associated to alternative approach presented in this paper.}
\end{figure}

\section{Deutsch's algorithm with inverse quantum engineering}

The Deutsch's algorithm is a quantum algorithm used to solve the following problem: \textit{Given a function $f(x):\{0,1\}\rightarrow \{0,1\}$, where $f(x)$ is promised to be \textit{constant} or \textit{balanced}. How can we show if $f(x)$ is constant or balanced?}
In 1980's, David Deutsch proposed an quantum algorithm to solve this problem \cite{Deutsch:85}, called \textit{Deutsch's algorithm}. The Deutsch's algorithm can be implemented by using a quantum circuit composed by three (or four, optional) Hadamard gates and an oracle $\Ocal$ that satisfies $\Ocal \ket{n}\ket{m} = \ket{n}\ket{n \oplus f(m)}$, as shown in Fig. \ref{DeutschCirc}. In addition, we need two qubits: the \textit{register} qubit, that will be read after circuit action, and an \textit{auxiliary} qubit, that can be discarded. 

As we said previous, we are interested to show how we can use HIQE for implementing the Deutsch's algorithm. Different from Ref. \cite{Santos:18-a}, here we will not provide Hamiltonians to simulate the quantum gates of the circuit in Fig. \ref{DeutschCirc}. We are interested to consider a protocol in which the Deutsch's algorithm can be implemented through an alternative approach. As a first consequence of the our approach, as shown in Fig. \ref{AlternDeutschCirc}, our scheme is composed by a single-qubit instead two ones. We can think about others approach where we could implement the Deutsch's algorithm using a single-qubit, for example, adiabatic quantum Deutsch's algorithm \cite{Sarandy:05-2}. Let us describe how our protocol works.

Without loss of generality, we consider that the qubit used in our scheme is initialized in state $\ket{0}$ (eigenstate of the $\sigma_{z}$ Pauli operator with eigenvalue $+1$). So, we implement an Hadamard gate for obtaining $\ket{+} = (\ket{0}+\ket{1})/\sqrt{2}$. In this step, the Hadamard gate is implemented by using the Hamiltonian in Eq. \eqref{GenericH}, where the simplest Hamiltonian for such operation is written as \cite{Santos:18-a}
\begin{eqnarray}
H_{\mathrm{Had}}(t) = \frac{\dot{\varphi}(t)}{2\sqrt{2}} \sigma_{z} + \frac{\dot{\varphi}(t)}{2\sqrt{2}} \sigma_{x} \mathrm{ , }
\end{eqnarray}
where $\varphi(t)$ satisfies $\varphi(\tau) = \pi$. The above Hamiltonian is a Landau-Zener type Hamiltonian and it can be experimentally projected by using quantum dots \cite{Fujisawa:09}, trapped ion \cite{Cui:16} or nuclear magnetic resonance \cite{Nielsen:Book}, for example.

Once we are using a different approach of the Deutsch's algorithm, here we need to define another oracle. In particular we will define the oracle as in Refs. \cite{Sarandy:05-2,Collins:98}, where we have $\Ocal^{\prime}\ket{n} = (-1)^{f(n)}\ket{n}$. The evolution operator $U_{\Ocal^{\prime}}(t)$ used to provide the correct output associate to oracle $\Ocal^{\prime}$ is given by Eq. \eqref{Uoperator}, with the vectors given by Eq. \eqref{phiState}. The initial state of this second step of the protocol is $\ket{+}$, so that the evolved state $\ket{\psi_{2}(t)} = U_{\Ocal^{\prime}}(t)\ket{+}$ will be
\begin{eqnarray}
\ket{\psi_{2}(t)} &=& \frac{1}{2\sqrt{2}} \left[ e^{i\varphi_{1}}+e^{i\varphi_{2}} - \left( e^{i\varphi_{1}} - e^{i\varphi_{2}}\right) \left( \cos \theta + e^{i\Omega}\sin \theta \right)\right] \ket{0} \nonumber \\
 &+& \frac{1}{2\sqrt{2}} \left[ e^{i\varphi_{1}}+e^{i\varphi_{2}} + \left( e^{i\varphi_{1}} - e^{i\varphi_{2}}\right) \left( e^{i\Omega}\cos \theta - \sin \theta \right)\right] \ket{1} \mathrm{ , }
\end{eqnarray}
therefore, it is easy to show that if we choose the parameters $\Omega(t)$ and $\theta(t)$ so that $\Omega(\tau)=0$ and $\theta(\tau) = \pi$, the output can be written as
\begin{eqnarray}
\ket{\psi_{2}(\tau)} &=& \frac{1}{\sqrt{2}} \left[e^{i\varphi_{1}(\tau)} \ket{0} + e^{i\varphi_{2}(\tau)} \ket{1}\right] \mathrm{ , }
\end{eqnarray}
where we can use $\varphi_{1}(t)$ and $\varphi_{2}(t)$ to encode the function $f:\{0,1\} \rightarrow \{0,1\}$ as $\varphi_{1}(\tau) = \pi f(0)$ and $\varphi_{2}(\tau) = \pi f(1)$. Now, by using that $e^{i \pi f(n)} = (-1)^{f(n)}$, we can write
\begin{eqnarray}
\ket{\psi_{2}(\tau)} &=& \frac{1}{\sqrt{2}} \left[(-1)^{f(0)} \ket{0} + (-1)^{f(1)} \ket{1}\right] \mathrm{ . }
\end{eqnarray}

We can note that $\ket{\psi_{2}(\tau)}$ is exactly $\Ocal^{\prime}\ket{+}$. Now, we can study the Hamiltonian that implements this dynamics. We note that the parameters $\Omega(t)$, $\theta(t)$, $\varphi_{1}(t)$ and $\varphi_{2}(t)$ should satisfy some boundary conditions, but we have not any condition about their time-dependence. Hence, as previous discussed, we can use this fact to provide feasible Hamiltonians. Firstly, we choose $\varphi_{1}(t) = \pi f(0)$ and $\varphi_{2}(t) = \pi f(1)$, and from Eq. \eqref{HamiltGen} we get the oracle Hamiltonian $H_{\Ocal^{\prime}}(t)$ as in Eq. \eqref{GenericH} where
\begin{subequations}
	\begin{eqnarray}
	\omega_{x}(t) &=& 2 \sin ^{2} \frac{F\pi}{2} \left[ \dot{\Omega}(t) \cos \Omega(t) \sin \theta(t) \cos \theta(t) + \sin \Omega(t) \dot{\theta}(t)\right] \mathrm{ , }\\
	\omega_{y}(t) &=& 2 \sin ^{2} \frac{F\pi}{2} \left[ \cos \Omega(t) \dot{\theta}(t) - \dot{\Omega}(t) \sin \Omega(t) \sin \theta(t) \cos \theta(t)\right] \mathrm{ , }\\
	\omega_{z}(t) &=& 2 \dot{\Omega}(t) \sin ^{2} \frac{F\pi}{2} \sin ^{2} \theta(t) \mathrm{ , }
	\end{eqnarray}
\end{subequations}
where $F = (-1)^{f(0)} - (-1)^{f(1)}$. Therefore, we can adjust the functions $\Omega(t)$ and $\theta(t)$ in order to obtain the simplest Hamiltonian. For example, because the $\Omega(t)$ and $\theta(t)$ needs to satisfy $\Omega(\tau)=0$ and $\theta(\tau)=\pi$, we can put $\Omega(t)=0$ and $\theta(t)= \pi t/\tau$. In this case we get the time-independent Hamiltonian
\begin{eqnarray}
H_{\Ocal^{\prime}} = \frac{\hbar}{\tau}\sin ^{2} \frac{F\pi}{2} \sigma_{y} \mathrm{ . }
\end{eqnarray}

It is important to highlight the role of $F$ above. Note that if we have a constant function, so $F = 0$, hence $H_{\Ocal^{\prime}}=0$. But this is not a problem of the theory, it is a trivial result of the protocol. In fact, since the input state of the second step is $\ket{+}$, an oracle associated with a constant $f$ can be simulated without any dynamics. It is important to mention that the information about $f$ should be encoded in the Hamiltonian. In addition, such result is not a particular characteristic of our approach, it is also present in adiabatic version of the Deutsch's algorithm \cite{Sarandy:05-2}.

To discuss about the last step of the protocol, we need to choose basis in which we will perform the measurement. If we want to measure the system in computational basis $\{\ket{0}, \ket{1}\}$, we need to apply a Hadamard gate. If we will measure the state in $\sigma_x$ basis, $\ket{\pm} = (\ket{0} \pm \ket{1})/\sqrt{2}$, no additional Hadamard gate need to be applied. In fact, let us consider the measurement in basis $\ket{\pm}$, by rewriting $\ket{\psi_{2}(\tau)}$ in such basis, we get
\begin{eqnarray}
\ket{\psi_{2}(\tau)} &=& \frac{(-1)^{f(0)} + (-1)^{f(1)}}{2} \ket{+} + \frac{(-1)^{f(0)} - (-1)^{f(1)}}{2} \ket{-} \mathrm{ . }
\end{eqnarray}
where the result is $\ket{+}$ if $f$ is constant, otherwise the result is $\ket{-}$.

\section{Search algorithm with inverse quantum engineering}

To provide a more practical example of a quantum algorithm that can be implemented with this approach, in this section we are interested to provide Hamiltonians for implementing the search algorithm. This algorithm was devised by Lov Grover in 1990's \cite{Grover:96,Grover:97}, where the problem solved was: \textit{given an disordered database with $N$ entires, one marked element $\ket{m}$ can be efficiently found (high probability) by using quantum mechanics}. In his paper, Grover considered an circuit composed by Hadamard gates and an oracle. Here we will make a different approach, where we will present Hamiltonians able to simulate such circuit. However, a detailed and good discussion about the original proposal of Grover's algorithm (search algorithm) can be found in Ref. \cite{Nielsen:Book}.

In general, we can consider a input state for the Grover's algorithm as an $n$-qubit state $\ket{0}^{\otimes n}=\ket{0}_{1}\ket{0}_{2}\cdots \ket{0}_{n}$. Thus, the first step is creating a \textit{uniform} distribution of all element of the disordered list, where we apply the Hadamard gate to each qubit and we get $\ket{\psi}=\ket{+}^{\otimes n}$. It is common we represent $\ket{\psi}$ in decimal basis $\ket{k}=\{\ket{0},\ket{1},\cdots,\ket{N-1}\}$, where $N = 2^{n}$ and each state $\ket{k}$ represents $\ket{0}=\ket{0}^{\otimes n}$, $\ket{1}=\ket{0}^{\otimes n-1}\ket{1}$, $\ket{2}=\ket{0}^{\otimes n-2}\ket{1}\ket{0}$, $\ket{3}=\ket{0}^{\otimes n-2}\ket{1}\ket{1}$ and so on. Therefore, the state $\ket{\psi}$ is written as
\begin{eqnarray}
\ket{\psi_{\mathrm{inp}}} = \frac{1}{\sqrt{N}} \sum_{k=0}^{N-1} \ket{k} \mathrm{ . } \label{PsiGro}
\end{eqnarray}

\begin{figure}
	\centering
	\includegraphics[scale=0.43]{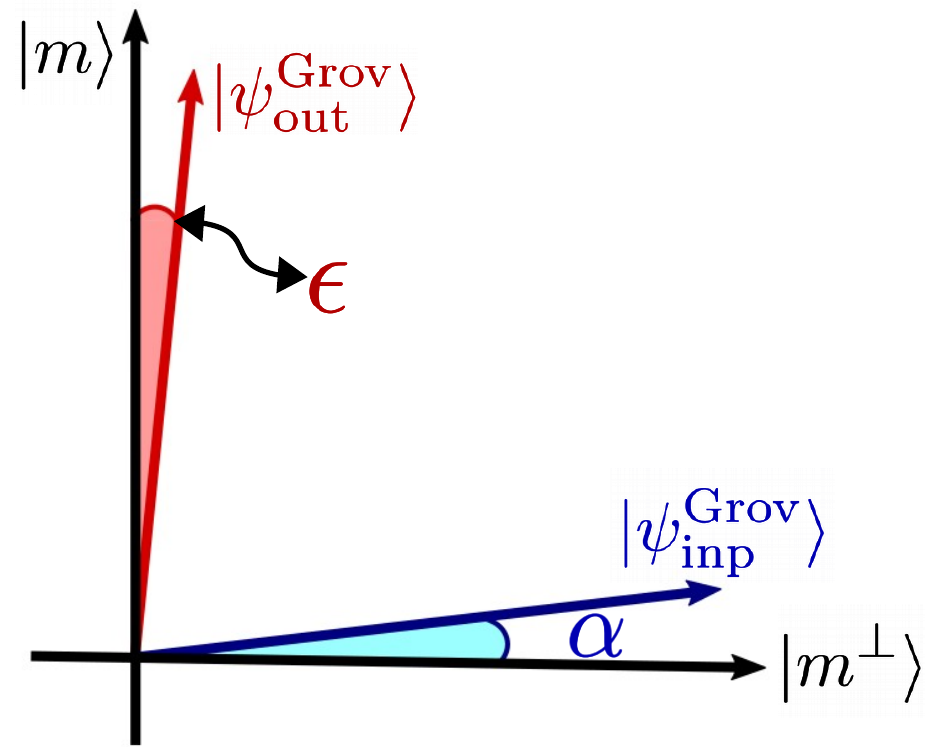}
	\caption{Geometrical representation of the bi-dimensional Grover's algorithm mapping.}
	\label{GroverFig}
\end{figure}

From this representation, we can map our $n$-qubit system into a hypothetic single-qubit system. Such mapping provide us a simple way to treat our study and it is used in others situations \cite{Nielsen:Book,Oh:14,Li:18}. Based on this representation, we can write the state in Eq. \eqref{PsiGro} as
\begin{eqnarray}
\ket{\psi_{\mathrm{inp}}^{\mathrm{Grov}}} = \frac{\sqrt{N-1}}{\sqrt{N}} \ket{m^{\perp}} + \frac{1}{\sqrt{N}} \ket{m} \mathrm{ . } \label{PsiGroTwo}
\end{eqnarray}
where we define the marked state $\ket{m}$ and $\ket{m^{\perp}}$, with $\ket{m^{\perp}}$ being composed by a uniform combination of all unmarked state, i.e.,  $\ket{m^{\perp}} = (1/\sqrt{N-1})\sum_{m \neq k} \ket{k}$. Thus, if we perform a measurement on the system, the probability $p_{m}$ of obtaining $\ket{m}$ is $p_{m}=1/N$, so that for $N \gg 1$, we have $p_{m} \ll 1$. To obtain an efficient protocol we need to drive $\ket{\psi_{\mathrm{inp}}^{\mathrm{Grov}}}$ to another state $\ket{\psi_{\mathrm{out}}}$ in which $p^{\mathrm{out}}_{m} \approx 1$.

To give a geometric representation of how our scheme works, consider the Fig. \ref{GroverFig}. We define the parameter $\alpha$ such that $\cos \alpha = \sqrt{(N-1)/N}$, in this case we get
\begin{eqnarray}
\ket{\psi_{\mathrm{inp}}^{\mathrm{Grov}}} = \cos \alpha \ket{m^{\perp}} + \sin \alpha \ket{m} \mathrm{ , } \label{PsiGrovtrig}
\end{eqnarray}
From Fig. \ref{GroverFig} we can note that if we want to obtain an output state $\ket{\psi_{\mathrm{out}}^{\mathrm{Grov}}}$ with $p^{\mathrm{out}}_{m} > p_{m}$, we should drive the system from $\ket{\psi_{\mathrm{inp}}^{\mathrm{Grov}}}$ to
\begin{eqnarray}
\ket{\psi_{\mathrm{out}}^{\mathrm{Grov}}} = \cos \alpha^{\mathrm{out}} \ket{m^{\perp}} + \sin \alpha^{\mathrm{out}} \ket{m} \mathrm{ , } \label{PsiOut}
\end{eqnarray}
where $\alpha^{\mathrm{out}} > \alpha$. From definition of the parameter $\alpha$ in Eq. \eqref{PsiGrovtrig}, we can see that $\alpha \approx 0$, therefore, for getting $p^{\mathrm{out}}_{m} \approx 1$, we should be able to achieve $\alpha^{\mathrm{out}} \approx \pi/2$.

We can show that our approach allow us to achieve this task by using the evolution operator $U(t)$ given in Eq. \eqref{Uoperator}. In fact, by writing $U(t)$ in basis $\{\ket{m},\ket{m^{\perp}}\}$ with
\begin{subequations}
	\begin{eqnarray}
	\ket{\phi_{1}(t)} &=& \cos [\theta (t)/2] \ket{m^{\perp}} + e^{i\Omega (t)} \sin [\theta (t)/2] \ket{m} \mathrm{ , } \label{phi1Gro} \\
	\ket{\phi_{2}(t)} &=& - \sin [\theta (t)/2] \ket{m^{\perp}} + e^{i\Omega (t)} \cos [\theta (t)/2] \ket{m} \mathrm{ , } \label{phi2Gro}
	\end{eqnarray} \label{phiStateGro}
\end{subequations}
and by choosing $\varphi_{1}(t) = 0$, $\varphi_{2}(t) = \varphi(t)$ and $\Omega(t)=0$, we get the evolved state
\begin{eqnarray}
\ket{\psi^{\mathrm{Grov}}(t)} &=& \frac{1}{2} \left[ (1+e^{i\varphi(t)})\cos \alpha + (1-e^{i\varphi(t)})\cos [\alpha-\theta(t)] \right]\ket{m^{\perp}} \nonumber \\
&+&\frac{1}{2} \left[ (1+e^{i\varphi(t)})\sin \alpha - (1-e^{i\varphi(t)})\sin [\alpha-\theta(t)] \right]\ket{m} \mathrm{ . }
\end{eqnarray}

Remarkably, note that if we impose $\ket{\psi^{\mathrm{Grov}}(\tau)}= \ket{\psi_{\mathrm{out}}^{\mathrm{Grov}}}$, the parameter $\varphi(t)$ in above equation could be picked so that $\varphi(\tau) = \pi$, and the final state $\ket{\psi^{\mathrm{Grov}}(\tau)}$ is written as in Eq. \eqref{PsiOut}, where $\alpha^{\mathrm{out}} = \alpha-\theta(\tau)$. To end, by computing the probability $p^{\mathrm{out}}_{m}$ we find $p^{\mathrm{out}}_{m} = \sin^{2}[\alpha-\theta(\tau)]$. Our result shows that there are infinity choices of $\theta(\tau)$ where $p^{\mathrm{out}}_{m} \approx 1$. More specifically, by imposing $\sin^{2}[\alpha-\theta(\tau)] \approx 1$, we find
\begin{eqnarray}
\theta(\tau) \approx (n+1/2)\pi + \alpha = \left(a+\frac{1}{2}\right)\pi + \arccos \left[ \sqrt{(N-1)/N} \right] \mathrm{ , } \label{BounConThe}
\end{eqnarray}
for any integer $a$. Moreover, in limit $N \rightarrow \infty$ we have $\theta(\tau) \rightarrow \left(a+1/2\right)\pi$, where $\theta(\tau)$, as well as $\theta(t)$, is independent on the number of elements of the database. This result shows that we are able to implement the Grover algorithm with an arbitrary probability $1-\epsilon^2$ from a careful choice of the parameter $\theta(\tau)$. In fact, by taking $p^{\mathrm{out}}_{m}$ around $\alpha-\theta(\tau) \approx\pi/2$, we get $p^{\mathrm{out}}_{m} = 1- \left[\alpha-\theta(\tau) - \pi/2\right]^2$, where we can identify $\epsilon = \alpha-\theta(\tau) - \pi/2$.

To find the Hamiltonian, we start from Eq. \eqref{HamiltGen}. We can show that, in basis $\{\ket{m},\ket{m^{\perp}}\}$, the Hamiltonian is written as in Eq. \eqref{GenericH} with
\begin{subequations}
	\begin{eqnarray}
	\omega_{x}(t) &=& \dot{\varphi}(t)\sin \theta(t) - \dot{\theta}(t) \cos \theta(t) \sin \varphi(t) \mathrm{ , }\\
	\omega_{y}(t) &=& 2 \dot{\theta}(t) \sin^2 [\varphi(t)/2]\mathrm{ , }\\
	\omega_{z}(t) &=& \dot{\varphi}(t)\cos \theta(t) - \dot{\theta}(t) \sin \theta(t) \sin \varphi(t) \mathrm{ , }
	\end{eqnarray}
\end{subequations}
where $\theta(t)$ needs to satisfy the Eq. \eqref{BounConThe} and $\varphi(t)$ should satisfy $\varphi(0)=0$ and $\varphi(\tau) = \pi$. In particular, by putting $\theta(t)= \mathrm{cte}$ we obtain $\omega_{y}(t)=0$, but now we will not take into account any consideration.

\section{Conclusion}

In this paper we have considered the role of Hamiltonian inverse engineering when we wish to implement quantum algorithm. Since such approach is a robust protocol against systematic errors \cite{Santos:18-a}, such algorithm can be efficiently performed at finite time. Remarkably, as we showed, the Grover algorithm can be effectively implemented with arbitrary probability though a single quantum evolution. In addition, as it can be obtained from others schemes of Grover algorithm \cite{Sarandy:05-2,Collins:98}, no auxiliary qubits are required and we can use single qubit analysis (from two-dimensional Grover's algorithm version). Since the robustness of our protocol was carefully studied in the literature \cite{Santos:18-a}, we believe that our approach constitutes a robust scheme for providing high fidelity dynamics and successful implementations of the algorithm studied in this paper.

\section*{Acknowledgments}

We acknowledge financial support from the Brazilian agencies CNPq and Brazilian National Institute of Science and Technology for Quantum Information (INCT-IQ).

\end{document}